# Integrated Planning of Multi-energy Grids: Concepts and Challenges

Marwan Mostafa[1]*, Daniela Vorwerk[2], Johannes Heise[1], Alex Povel[3], Natalia Sanina[4], Davood Babazadeh[1], Christian Töbermann[4], Arne Speerforck[3], Christian Becker[1] and Detlef Schulz[2]

[1]Institute of Electrical Power and Energy Technology, TU Hamburg, Hamburg, Germany

[2] Electrical Power Systems, Helmut Schmidt University, Hamburg, Germany

[3] Institute of Engineering Thermodynamics, TU Hamburg, Hamburg, Germany

[4]Technische Hochschule Lübeck, Lübeck, Germany

* marwan.mostafa@tuhh.de

*Abstract* - **In order to meet ever-stricter climate targets and achieve the eventual decarbonization of the energy supply of German industrial metropolises, the focus is on gradually phasing out nuclear power, then coal and gas combined with the increased use of renewable energy sources and employing hydrogen as a clean energy carrier. While complete electrification of the energy supply of households and the transportation sector may be the ultimate goal, a transitional phase is necessary as such massive as well as rapid expansion of the electrical distribution grid is infeasible. Additionally, German industries have expressed their plans to use hydrogen as their primary strategy in meeting carbon targets. This poses challenges to the existing electrical, gas, and heating distribution grids. It becomes necessary to integrate the planning and developing procedures for these grids to maximize efficiencies and guarantee security of supply during the transition. The aim of this paper is thus to highlight those challenges and present novel concepts for the integrated planning of the three grids as one multi-energy grid.**

## I. Motivation

In 2021 the city of Hamburg voted on a new legally binding climate plan, placing stricter and lower carbon emission targets, and moving the carbon neutrality target from 80% carbon emission reduction compared to 1990 by the year 2050 to 95% carbon emission reduction by the year 2045 [1]. Achieving these climate targets with electrification alone is not feasible due to the lack of capacities in the electrical distribution grid and the short time frame to increase the capacity of the electrical grid, especially with the high industrial energy demand and limited area to generate renewable energy locally due to the city's urban and densely built character and very rare wind priority areas. The research project Integrated Grid Development Planning, iNeP (in German "**I**ntegrierte **Ne**tzentwicklungs-**p**lanung") aims to develop a general approach and methodology for integrating the planning processes of the three individual energy grids, namely electricity, gas and heating, into one process in form of road map for grid expansion.

The main challenges facing integrated grid planning are to name a few: the lack of data availability, an absence of common scenario framework, proliferation of sector coupling technologies, new energy generation technologies, energy security, and also various characteristics and dependencies of time and temperature for the consumers of the three energy carriers until today. Sector coupling technologies offer especially complex challenges to the status-quo of demand forecast as well as grid planning methods and processes.

This paper presents different concepts to overcome these challenges and enable the integrated planning of the gas, heating, and electrical grids in light of the necessity of sector coupling to achieve the climate targets.

## II. Grid Planning – State of the Art

In this section, the state-of-the-art in the planning of energy grids are briefly described and the requirements and thoughts for future approaches are mentioned.

### A. Demand prediction

Conventionally, suppliers and consumers of energy are distinct, spatially separate parties, necessitating transmission grids and distribution grids respectively. Should temporal distinctions be present as well, storage capacities become necessary. In addition, supply follows demand in forecasting grid as well as supply expansion needs. Currently, demand determination is coarse, relying on inflexible known past values, and not integrated with other sectors.

In the future, energy demand has to become substantially more dynamic. In fact, wherever feasible, demand might start tracking supply. Further, countless, partly yet to be established technologies will decouple useful and primary energy ties. Conventionally, only a handful of established avenues exist, coal usually supplies electricity, so do solar and wind, gas often covers heat demands. Eventually, solar and wind might produce gas in the first place, which in turn might be hydrogen or synthetic natural gas, either one capable of covering a wide gamut of green useful energies.

### B. Conventional Grid planning methods

In general, the urban network comprises three different energy grids (electrical grid, gas grid and heat grid).

#### 1) Electrical Grid

The planning principles of the VDE Forum Netztechnik/

Netzbetrieb (FNN) [2] are applied to the planning of electricity grids in Germany. Network planning considers both the maximum power of the consumers and the maximum transmission capacity of the operating equipment.

   *2) Gas Grid*

   The gas network applies the planning principles and rules of the German Technical and Scientific Association for Gas and Water (DVGW) [3]. Annual volumes and standard temperature dependent load profiles are used to determine the demand on the basis of which the gas network has been designed to day.

   *3) Heating Grid*

   In the district heating network, the planning principles of the "Energy Efficiency Association for Heating, Cooling and CHP" (AGFW) [4] apply. Since there used to be little flexibility in the distribution and use of energy, the planning of the networks was based exclusively on the prediction of static demands [5].

   III. NOVEL PLANNING CONCEPTS AND APPROACHES FOR MULTI-ENERGY GRIDS

   The overall concept of integrated grid planning proposed in this paper is shown in **Figure 1**. The scenarios and frameworks are part of the integral planning since cross-sector thinking is already required here. The next step of integrated planning is the modeling of the multi energy grid. Finally, the network expansion planning is supported by appropriate tools for optimization.

   *A. Neutral cell areas*

   When it comes to an integrated approach including demand prediction as well as grid planning for three different kinds of energy carriers, the definition of spatially delimited neutral cell areas might be wise. They serve as a basis for regionalized description in an arbitrary resolution for the consideration on the level of distribution grids and should be chosen independently of the specific energy carriers. They overcome conventional standard values for grid planning of electric, gas or heating grids. Their characterization shall not only include energetical aspects and standard empirical values for grid planning of the three energy carriers as e. g. presented in [6]–[9] but also take other aspects into account for improved prediction of future urban and demographic developments.

   *1) Characterization of neutral cell areas*

   For the characterization of neutral cell areas to name a few, the following topics and corresponding exemplary aspects as given in **Table 1** shall be regarded.

   Describing the aspects, specific numerical key factors can be developed. Formulating factors as numerical values permits the comparability between the cell areas on one hand and the expression of developments and interdependencies as mathematical correlations on the other hand, for example, numerical values such as protected areas or number of buildings seem to be rationally related to the neutral cell's area. Installed power generation or storage and grid capacities can be set in a ratio to the prevailing demands, whereas other key factors are developed by setting e.g., the number of private cars in reference to the number of inhabitants. They give an idea about the changeability of one area and how specific trends will develop in a regionalized way. This should be analyzed in detail, which values might stay constant and stable over time, which might change fast and how strong the interdependency between the key factors is, e. g. using statistical correlation tests [10]. Using techniques like the scenario technique [11]–[13] the interdependency can be considered to make future estimations. It also has to be taken into account, which key factors are controllable and can be affected in a positive way within the approach of integrated grid planning, and which only depend on outside effects.

   In this approach, all the characteristics of neutral cell areas are considered homogeneous within the cell. Also, the high relevance of neighbored zones and their characteristics shall be mentioned here because the choice of the cells' borders are chosen arbitrarily, as described in the following section.

   *2) Geographic design of neutral cell areas*

   In the following, three possible shapes of neutral cell areas shall be introduced briefly: City districts, raster, and floating cells, as pictured in **Figure 2**.

   As a first approach, individual city districts shall be regarded, because detailed statistics for urban, demographic, and socio-economic aspects regarding the past and predictions regarding future developments are available on that basis [14]. Assuming every city district as more or less best described using its average parameters has weaknesses regarding large districts with high inhomogeneous structure. Using equidistant raster as the second approach for neutral cell areas, that disadvantage can be conquered. In opposite to accessible GIS-Data taken from [15], which can be easily included in both kinds of neutral cell areas via geographic assignment, other data from the district statistics cannot be transferred directly, so suitable distribution keys, e. g. share of buildings, have to be found. Using districts as well as raster, the geographic limits are fixed from the beginning and their content is regarded afterwards to develop key factors. In opposite to that, the approach of "floating cells" shall be able to define the cell areas' limits dependent on reasonable homogenous structures e.g., settlements, protection areas, industrial or business areas etc. The development of that floating zones is a huge challenge, because suitable aspects have to be chosen to serve as the main dictating aspects, and the algorithm needs detailed instruction about how a cell area's border is explicitly defined. Another challenge occurs in terms of size and simultaneity. The smaller such a cell area becomes, the more difficult it becomes to make reliable statements for its development and simultaneity factors.

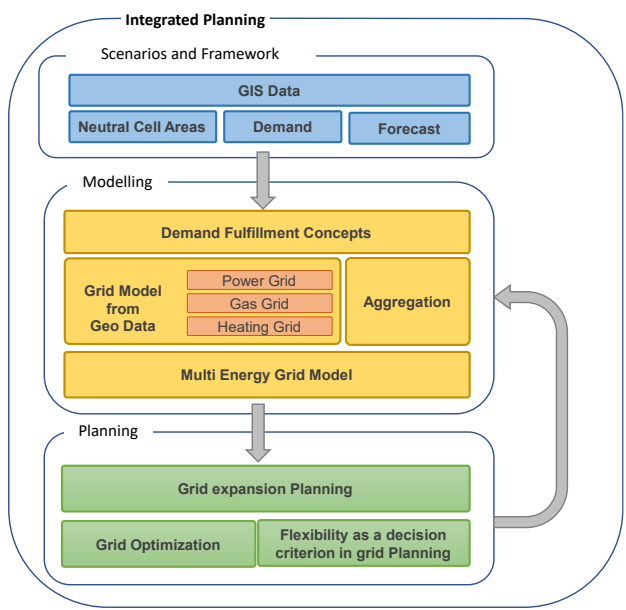

**Figure 1** Overall concept of integrated planning.

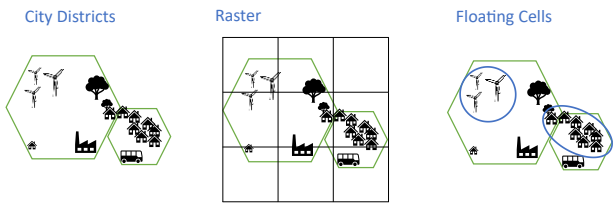

**Figure 2** Approaches for geographic design of neutral cell areas.

*3) Demand fulfillment concepts of neutral cell areas*

A demand fulfillment concept of a holistic energy system in an integrated way can be of high diversity. It includes various technical end devices, but also centralized or decentralized storages, power generation or supply from the transport level, coupling devices between the different grids like Power-to-Heat, Power-to-Gas, or other technologies. Regarding the mobility sector, the fulfillment concept e.g., includes private or public transport.

The full separation of demand fulfillment from the demand itself enables a high number of degrees of freedom within the design of a fulfillment concept. Some thoughts should be made in the very beginning to restrict the solution space in the first view:

- The main boundary condition of climate neutrality restricts the existence of some technologies no later than in 2045.
- Some concepts of absolute extrema should be developed to get an idea of the probable limits.
- Some technologies compete with each other, and their redundancy might be not useful or even impossible.
- The initial state of the demand fulfillment concept as given from today's key factors should be considered and the realistic feasibility of strong modifications within the prevailing energy system should be evaluated critically.
- The objective of the concept should be clarified.

The objective of the coverage concept can be quite variable. Besides the minimization of primary energy input, high self-sufficiency of the individual cell areas or minimal construction costs can be part of the investigation. For multi-objective approaches, the single objectives can be weighted, and the optimization can be executed using heuristic approaches such as Greedy etc. [16]. The developed initial key factors give first approaches for fulfillment concepts within the individual cell areas and possibly on their basis some solutions can be excluded. For example, the operation of heating grids is only economical at a specific level of heating density. Individual local heating grids might take place close to local industry with dense building structure around. Also, areas with high number of buildings might be suitable for future expansion of solar energy. The virtual construction of new technologies or changing the share of technologies in neutral cell areas will change their key factors over time and might have impact on others, which can be analyzed using the valid correlations based on the key factors.

*4) From fulfillment concept to grid planning*

Regarding a neutral cell area as one strictly limited area concentrating everything in it on one node, its overall power balance must be fulfilled at every timestep. For knowledge of the grid planning itself, not energies but the maximum power flows have to be known for design of transport capacities. To get from annual demands to necessary power delivery values for the specific energy carrier, electric power, gas or heat, the time dependency of the demand must be considered. That can be realized considering standard load profiles [17], [18] or temperature depended on load profiles (e. g. presented in [19] for heat-pumps) in the first approach. The power to deliver from the specific energy carriers can be estimated using efficiencies or COPs of the prevailing technologies. Their shares are also a character of the neutral cell area (see **Table 1**). The timepoint of maximum power delivery into one cell area gives an idea about the necessary transport capacity. That case does not obligatorily occur at the same time for all cell areas and highly depends on the cells' specific demand fulfillment concept. For improved approaches, it can be investigated, how flexibilities may support defusing maximum power delivery (see section E).

**Table 1** Fields and aspects for characterization of neutral cell areas.

| Topic | Aspects |
|---|---|
| Energy aspects | - Installed capacities of conventional plants, renewable resources, storages<br>- Heat density<br>- Potential for solar energy use |
| Socio-economic | - Demography, Aging structure<br>- Living space<br>- Income |
| Geography | - Base area<br>- Protected area, forests |
| Mobility and transport | - Car density<br>- Public transport stations<br>- Charging stations<br>- Bus depots<br>- Parking lots/Parking houses |
| Building and Consumer structures | - Number of Buildings<br>- Local supply, trade<br>- Industrial site |

*B. GIS data, demand & forecast*

Grid planning is an inherently spatial process, as hinted in II.A, and energy grid operators have long based their work on geographic information systems. Mutually beneficial, open tools like geopandas or QGIS as well as open standards such as OGC's GeoPackage may facilitate this. Alongside open government paradigms, allowing for open data provision like the city of Hamburg's ALKIS. When supplied with (open or internal) data, these tools allow vertically integrated pipelines to emerge, providing rich insights from raw data. Owing to open, established standards, these can then be scaled horizontally: ALKIS is a nationwide format. Provided as a GeoPackage, it could be parsed by anyone with ease. Pipelines constructed in the context of the city of Hamburg will work nationwide, with few adjustments

A part of one such vertical pipeline is the Hamburg *Wärmekataster* [20]. This digital heat cadaster contains annual heat consumption data for building blocks. It was reimplemented based on the original method outlined in [21]. Over the original, public version, it affords:

- individual building-level instead of building-blocks aggregated data (essentially undoing the previously introduced, artificial aggregation in the online version),
- feature enhancements:
  - power (energy flows) level computations,
  - integration of internal, corporate data,
- performance improvements and modernization, the latter mainly through contemporary data formats (GeoPackage, JSON etc.) as well as sources (web APIs of the city of Hamburg).

A building's overall heat losses may be modelled as a heat flow through a plane wall, given as

$$\dot{Q} = kA\Delta T = kA(T_{\text{outer}} - T_{\text{inner}}), \quad (1)$$

with $k$ as the thermal transmittance and $A$ as the building's enveloping area. While it is infeasible to determine either for each building, they may be assumed approximately constant. With the annual heat demand known, inner (room) temperatures assumed as a constant 20 °C [22] and outer, i.e. ambient, temperatures known from weather data [23], the factor $c = kA$ can be solved for via integration (summation).

Knowing $c$ and *daily* average ambient temperatures, *daily* average heat loss flows are known as well. Integrating (multiplying) these for a day's duration each, a final formulation in analogy to *degree days* [22] results as

$$Q_{\text{d}} = Q_{\text{a}} \frac{T_{\text{i}} - T_{\text{o}}}{\sum_{j=1}^{j=n} T_{\text{i}} - T_{\text{o},j}} \quad (2)$$

Applying this heuristic to each building in the Hamburg ALKIS, approximate daily heat consumptions for each building result. These can then be used as the input to subsequent demand fulfilment concepts.

Forecasts are handled differently, using Agent-Based Modelling within the Mesa (Python) framework [24]. Using the heat cadaster built in the previous step as the status quo at timestep zero, agents (households) are initialized across all of Hamburg. The main purpose of the modelling in existence so far is to track the development of heating technologies in use (gas and oil boilers, heat pumps, district heating, electric rod and more).

The agents are assigned a set of attributes (the model's parameters), like initial heating technologies (randomly initialized from known, aggregate data), income and expenditures, their willingness to switch technologies (essentially a culmination of social and political views and more), initial funds, saving quotas and more. In each time step, agents are subjected to changes in some these attributes (while others are constant throughout). As a result, they might switch their heating technologies, e.g., from gas boilers to heat pumps, if that decision is rational within their limited frame of reference. Hysteresis and inertia are used to model human irrationalities.

A main driver for the development of heating technologies across time steps are their costs. These can be adjusted per technology per time step. As such, scenarios such as political endorsement (realized through fiscal promotions) or technological breakthroughs (or even just steady progress) can be expressed as cost time series. Their effect is then visible in differing development of technologies over time. This approach is *bottom-up* and offers, among others, one core advantage: Suppose a legislator mandates or aims for, for example, 90% heat pump penetration in private households until 2045. Through a Monte-Carlo meta-analysis, many parameters sets and therefore scenarios can be examined and those in agreement with the target can then be extracted. Not only will these meet the target, the *path* is then also known, translating into actionable insights ("By 2030, implement a 40% price reduction."). While still simplistic, the model can be arbitrarily enriched with data and decision logic, leading to more precise analyses. The main challenge is that of acquiring input data, as logic can be added much more easily.

*C. Grid model from GIS data*

Due to the lack of GIS-data availability regarding the distribution grids, due to technical or mostly legal obstacles, an electricity grid for the city is to be generated for corresponding simulations as part of the research project. A first grid design could be generated with the Fraunhofer internal program "DaVe", where "DaVe" stands for

(energy) Data Analysis and Processing (in German "(Energie-) **Da**tenanalyse und -**Ve**rarbeitung") [25]. This program allows an automatic generation of network models based on open data.

For the creation of the network topology on the low voltage level, methodologies have been developed to derive it from the given geographical information [25]. To determine the house connection nodes, the centers of the building plots are taken from the OpenStreetMap (OSM) data. Then, suitable network connection points must be determined. They assume that the lines run along the streets and are derived to find the shortest distance to the house connection node in the nearest street. Based on the determined network nodes, suitable lines are then generated to connect the buildings with each other. In a first step, all building nodes are connected to the corresponding network connection points. In the next step, the network connection nodes are combined with each other via lines along existing streets. Furthermore, additional nodes are placed on the road intersections.

*D. Aggregation into a multi-energy grid model*

Graph theory based approaches to represent the energy grid allow for the combined representation of the multi-energy grid by representing sector-coupling technologies as vertices connecting multiple grids together, where the energy carrier is represented by an edge, and the energy flow direction is represented by the direction of the edge connecting the vertices. Ultimately the entire multi-energy grid can be represented as one graph as shown in **Figure 3**.

In addition to the methods of grid calculation for the individual sectors applied so far, there are already approaches for a cross-sector grid calculation [26] for district grids. Here, the method described in [27] must be mentioned, which allows to calculate coupled district energy systems. The coupling of the energy networks was mapped here using a cross-sector efficiency matrix. The now coupled load flow equations for electricity, heat and gas were solved using a Newton-Raphson approach. However, this method cannot be applied to larger energy systems, such as the distribution grid, since the aspect of the load variance on the network structure is not considered here. In distribution networks, however, the network structure must be adapted to a changing consumption to avoid future network bottlenecks. In [28], [29] methods for energy flow calculation of a multi-energy system are shown. These methods however do not consider hydrogen as an energy carrier and a coupling technology between the electrical and gas grids and must be expanded upon in this regard. This will be adapted and implemented in a graph theoretical model of the multi-energy grid to perform the load flow calculations.

In [30] the optimal location of electric vehicle chargers is determined using graph theory, a more advanced version of this can be used in our project to determine the optimal location of grid coupling points such as electrolysis plants regarding cost, and grid constraints.

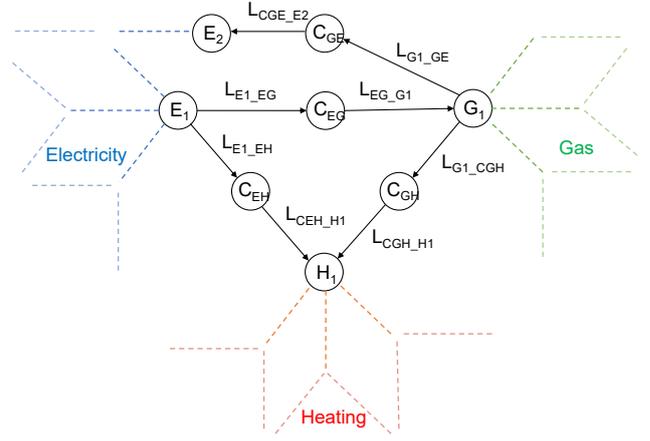

**Figure 3** Graph theoretical representation of the multi-energy grid.

Graph theory based approaches also offer great advantages in solving complex grid optimization problems. It enables applying energy routing algorithms as well as cost minimization algorithms, an application for this can be seen in [31].

*E. Flexibility as a decision criterion in grid planning*

As mentioned previously, energy demand will be more dynamic in the future. Therefore, flexibility will play a decisive role in future energy grids. With the help of flexibilities in the energy distribution networks, short-term network bottlenecks can be avoided. In addition, taking flexibility options into account in an early stage of the planning phase can lead to cost reductions in grid expansion. Sector coupling and storage facilities in particular offers opportunities to use flexibility across energy sectors.

The approach presented in the section "D" is to be extended to consider flexibility. The aim is to represent the flexibility provided by storage and coupling aggregates and to integrate it into the grid optimization. One advantage of this consideration is that the grid expansion planning can be directly compared to the development of coupling aggregates and storage units, e.g., the use of heat pumps that are operated in conjunction with a heat storage system to serve the grid. Due to the flexibility provided and the grid-serving operation of these coupling units, the grid expansion can be adapted accordingly. Another advantage is that the placement of storage units and coupling elements can also be optimized and improve the system resilience [32].

## IV. VALIDATION

An important stage of this research project is the validation of the results achieved using the novel methods. Test cases are to be constructed and calculated using the novel methods and serve as a basis for the industrial partners to calculate using their inhouse state of the art grid planning and calculation tools. These results will be then used to analyze the scale of their possible integration in the planning processes of the distribution grid operators. The aim of the project is to define a roadmap for the integrated grid expansion, which will serve as a guideline for lawmakers and grid operators in the city of Hamburg in their plans for grid development and expansion. Further cooperation with grid operators from other regions should also investigate the applicability of the methods discussed here to cities with different grid and load structures.

## V. Conclusion and Outlook

The paper mentions the challenges facing planning of distribution grids in light of decarbonization, distributed generation and demand prediction in a sector coupled energy system. It also presents novel concepts and methods to overcome such challenges, which together form a holistic approach for an integrated grid planning and development. The approach starts from characterizing the prevailing infrastructure as the initial basis, modelling the existing grid, demand prediction and fulfillment, load flow calculations to optimizing the grid and maximizing its potential. It also presents the possibility of utilizing flexibilities not implemented yet in current grid planning methods. The research project is currently in its starting phases, and the methods presented here provide a methodological approach for the plan the project is following.

An expansion to DaVe program that will add the same feature to gas grids in addition to the electrical grid is currently in development. Currently DaVe covers all voltage levels for Germany in the power sector and the high-pressure level for Europe in the gas sector. Furthermore, the basic grid elements can be generated.

The approach of neutral cell areas and the concept of their characterization by development of various key factors on different topics besides energetic aspects, socio-economic, geographical, mobility and transport and consumer structures has been addressed. They serve as a basis for and integrated and regionalized consideration of a whole industrial metropolis. Developments within that neutral cell areas get visible by the change of the specific key factors and their interdependencies. Regarding the demand fulfillment concepts, formulating some restrictions and boundary conditions are useful at the beginning to keep the problem lucid.

## Acknowledgment

This research project is supported by the German federal ministry of economic affairs and climate action (BMWK) under the agreement no. 03EWR007A-V.

The authors would like to thank the industrial project partners Gasnetz Hamburg GmbH, Stromnetz Hamburg GmbH and Hamburger Energiewerke GmbH for their support and insight in the processes and challenges in grid operation and planning.